\documentstyle[amssymb,aps,prb]{revtex}
\begin{document}
\title{Nature of bonding and electronic structure in MgB$_{2}$, a boron
intercalation superconductor}
\author{K.D. Belashchenko, M. van Schilfgaarde$^{\dagger}$ and V.P. Antropov}
\address{Ames Laboratory, Iowa State University, Ames, Iowa 50011\\
$^{\dagger }$Sandia National Laboratories, Livermore, California 94551}
\date{\today}


\twocolumn[\hsize\textwidth\columnwidth\hsize\csname@twocolumnfalse\endcsname

\maketitle

\begin{abstract}
Chemical bonding and electronic structure of MgB$_2$, a boron-based
newly discovered superconductor, is studied using self-consistent band structure techniques.
Analysis of the transformation of the band structure for the hypothetical series of
graphite -- primitive graphite -- primitive graphite-like boron -- intercalated boron,
shows that the band structure of MgB$_2$ is graphite-like, with $\pi$ bands falling deeper than
in ordinary graphite.  These bands possess a typically delocalized and metallic, as opposed to
covalent, character. The in-plane $\sigma$ bands retain their 2D covalent character,
but exhibit a metallic hole-type conductivity. The coexistence of 2D covalent in-plane and
3D metallic-type interlayer conducting bands is a peculiar feature of MgB$_2$. We analyze
the 2D and 3D features of the band structure of MgB$_2$ and related compounds, and their 
contributions to conductivity.
\end{abstract}

] 

The recent discovery of superconductivity (SC) in MgB$_{2}$\cite{aki} has
stimulated significant interest in this system\cite{mazin,budko}. One of the
first questions is whether MgB$_{2}$\ represents a new class of
superconductors, or it may be related to other known classes in terms of its
bonding and electronic properties. The crystal structure of MgB$_{2}$ may be
regarded as that of completely intercalated graphite\cite{Bur} with carbon
replaced by boron (its neighbor in the periodic table). In addition, MgB$%
_{2} $ is formally isoelectronic to graphite. Therefore, understanding the
bonding and electronic properties of MgB$_{2}$ and its relation to graphite
intercalation compounds (GIC), some of which also exhibit SC, is desirable
and is the subject of this paper.

The search for HTSC in carbon phases started in 1965 when T$_{c}$ = 0.55 K
in KC$_{8}$ was reported, and SC was subsequently explained in terms of the
interaction of $\pi $-derived bands with the longitudinal phonon modes of
the graphite layer\cite{REVIEW}. Similar conclusions for the electronic
structure at the Fermi level were derived for another GIC, LiC$_{6}$. The
highest T$_{c}$ achieved for a GIC was 5 K\cite{belash}. A parallel
development was that of SC in Bechgard salts, the organic charge-transfer
systems. Within this family of organic materials exhibiting carbon $\pi $%
-band conductivity, T$_{c}$ has been raised to 12.5 K\cite{12.5}. It should
be noted that the majority of these compounds are unstable at normal
conditions and high-pressure experiments are most common. This part of GIC
studies was reviewed in Ref.~\onlinecite{REVIEW}.

Later research shifted to the area of carbides and nitrides of transition
metals where metallicity and conductivity are mostly due to transition metal
atoms. Only in 1991 was it shown that alkali-doped C$_{60}$ exhibits SC with
maximum T$_{c}$ of 33 K\cite{Tanigaki}. It was the first representative of a
3D carbon-based metal with modified (compared to graphite) coupling of
lattice vibrations to the conduction electrons due to the curvature of the
molecule but still with $\pi $-band derived states at the Fermi level.

We have calculated the electronic structure of MgB$_{2}$ and some related
compounds using the Stuttgart LMTO-TB (ASA) code and full-potential LMTO
(FLMTO) method\cite{nfpmethod}. A local exchange-correlation potential has
been used in both cases (addition of gradient corrections did not change the
results). It appears that a general analysis of energy bands may be safely
done in ASA, while reliable treatment of charge densities and anisotropy of
transport properties of these compounds requires more accurate FLMTO
calculations.

In order to examine the relation between the band structure of MgB$_{2}$ and
that of graphite we studied the following hypothetical sequence of
intermediate materials: carbon in the `primitive graphite' (PG) lattice with
no displacement between layers as in MgB$_{2}$ and in GICs, using graphite
lattice parameters; boron in the PG lattice with $a$ as in MgB$_{2}$ and $%
c/a $ as in graphite; boron in the PG lattice with $a$ and $c/a$ as in MgB$%
_{2}$; LiB$_{2}$ in the same structure; MgB$_{2}$ itself. The results of
some of these calculations are shown in Fig.~1. The band structure of PG
carbon shown in Fig.~1a is very similar to that of graphite (see e.g. Ref.~%
\onlinecite{Freeman} and references therein) with the appropriate
zone-folding for a smaller unit cell. (This is quite natural because of the
weak interlayer interaction.) Boron in the same lattice but scaled to match
the MgB$_{2}$ in-plane lattice parameter (not shown) has nearly identical
bands with the energies scaled by the inverse square of the lattice
parameter, in agreement with common tight-binding considerations\cite%
{Harrison}. Fig.~1b shows the natural enhancement of the out-of-plane
dispersion of the $\pi $ bands when the interlayer distance is reduced.
Figs. 1c and 1d demonstrate that `intercalation' of boron by Li or Mg
produces a significant distortion of the band structure, so that the role of
the intercalant is not simply one of donating electrons to boron's bands
(which would return the band structure to that of PG carbon shown in
Fig.~1a). The main change upon intercalation is the downward shift of the $%
\pi $ bands compared to $\sigma $ bands. For Li this shift of $\thicksim $%
1.5 eV is almost uniform throughout the Brillouin zone. Replacement of Li by
Mg shifts the $\pi $ bands further, but this shift is strongly asymmetric
increasing from $\thicksim $0.6 eV at the $\Gamma $ point to $\thicksim $2.6
eV at the A point. In addition, the out-of-plane dispersion of the $\sigma $
bands is also significantly enhanced. In LiB$_{2}$ the filling of the
antibonding $\sigma $ band is nearly the same as in PG boron, while in MgB$%
_{2}$ the Fermi level shifts closer to the top of the antibonding $\sigma $
band.

The Fermi surface (FS) for MgB$_{2}$ obtained in the FLMTO method is
practically identical to Fig.~3 of Ref.~\onlinecite{mazin}. The incompletely
filled antibonding $\sigma $ band with predominantly boron $p_{xy}$
character forms two hole-type cylinders around the $\Gamma -{\rm A}$ line,
while the $\pi $ bands form two planar honeycomb tubular networks: one
electron-type one centered at $k_{z}=0$ and another similar, but more
compact, hole-type one centered at $k_{z}=\pi /c$. The lowering of the $\pi $
band in MgB$_{2}$ compared to PG boron is connected with charge accumulation
in the interstitial region. This lowering is greater at the A point compared
to the $\Gamma $ point, because the interstitial charges favor antisymmetric
overlap of the boron's $p_{z}$ tails.

The nature of bonding in MgB$_{2}$ may be understood from the charge density
(CD) plots shown in Fig.~\ref{charge}. As seen in Fig.~\ref{charge}a,
bonding in the boron layer is typically covalent. The CD of the boron atom
is strongly aspherical, and the directional bonds with high CD are clearly
seen (see also Ref.~\onlinecite{Medv}). It is worth noting that the CD
distribution in the boron layer is very similar to that in the carbon layer
of graphite (see e.g. Ref.~\onlinecite{Freeman}). This directional in-plane
bonding is also obvious from Fig.~\ref{charge}b showing the CD in a cross
section containing both Mg and B atoms. However, Fig.~\ref{charge}b also
shows that a large amount of valence charge does not participate in any
covalent bonding, but rather is distributed more or less homogeneously over
the whole crystal. Further, Fig.~\ref{charge}c shows the difference of the
CD of MgB$_{2}$ and that of hypothetical NaB$_{2}$ in exactly the same
lattice. Not only does it show that one extra valence electron is not
absorbed by boron atoms but is rather delocalized in the interstitials; it
also shows that some charge moves outward from boron atoms and covalent
in-plane B-B bonds. Fig.~\ref{charge}d shows the CD difference between the
isoelectronic compounds MgB$_{2}$ and PG carbon (C$_{2}${}). In MgB$_{2}$,
the electrons see approximately the same external potential as in C$_{2}$,
except that one proton is pulled from each C nucleus and put at the Mg site.
It is evident that the change C$_{2}${}$\rightarrow $MgB$_{2}$ weakens the
two-center $\sigma $ bonds (the charge between the atoms is depleted) and
redistributes it into a delocalized, metallic density.

Thus, the structure of MgB$_{2}$ is held together by strongly covalent
bonding in boron sheets and by delocalized, `metallic-type' bonding between
these sheets. A peculiar feature of this compound is that electrons
participating in both of these bond types provide comparative contributions
to the density of states (DOS) at the Fermi level (see below). This
distinguishes MgB$_{2}$ from closely related GICs where covalent bonds in
the carbon sheets are always completely filled, while the nearly cylindrical
parts of the FS commonly found in those compounds are formed by
carbon-derived $\pi $ bands which are also responsible for conductivity in
pristine graphite\cite{REVIEW}.

Because of the coexistence of two different types of chemical bonds, it is
desirable to find the contributions to the total DOS and transport
properties from separate pieces of the FS originating from 2D covalent and
3D metallic-type bonding. Such decomposition is shown in Fig.~\ref%
{decomposition} for the total DOS and for the in-plane ($xx$) and
out-of-plane ($zz$) components of the tensor $\sigma_{\alpha \beta }=\int
v_{\alpha }v_{\beta }\delta (\varepsilon \left( {\bf k}\right) -E_{F})d^{3}%
{\bf k}$, with $v_{\alpha }$ being a component of the electronic velocity.
One can see that the 3D (metallic-type bonding) and cylindrical (covalent
bonding) parts of the FS contribute, respectively, about 55\% and 45\% to $%
N(0)$. $N(\varepsilon )$ for the hole-type zones rapidly decreases with
increasing $\varepsilon $ and already at $\varepsilon \thickapprox 0.8$ eV
above $E_{F}$ the total DOS is almost completely determined by the 3D
electron-type band. The latter contribution is almost constant and probably
is not related to the change of SC properties under pressure or with doping.
The corresponding contribution to $\sigma$ exceeds all other
contributions (more than 50\% for $\sigma _{zz}$ and $\sigma _{xx}$) and is
virtually isotropic. Holes in the in-plane B-B covalent bands (two cylinders
at the FS) as expected have clearly anisotropic behavior contributing about
30\% to $\sigma _{xx}$ and virtually nothing to $\sigma _{zz}$. The 3D hole-type
part of the FS is also notably anisotropic with predominantly
$z$-axis conductivity. The total $\sigma$ has a relatively small anisotropy
at $E_F$ with $\sigma_{xx}=2.66$
and $\sigma_{zz}=2.18$ $\rm{Ry}\cdot(a/2\pi)^2$.

The main challenge is to suggest another possible SC with a higher T$_{c}$.
We studied several similar systems which, as we believe, may be interesting
for experimental studies. Several isoelectronic compounds demonstrate very
different behavior. Hypothetical BeB$_{2}$ and ZnB$_{2}$ in MgB$_{2}$
structure are very similar to MgB$_{2}$ in terms of the electronic
properties. Our ASA calculations of these compounds with lattice parameters
of MgB$_{2}$ produced very similar band structures with nearly identical FS.
Presumably a smaller radius of the Be ion and a larger one of the Zn ion
will generate some difference but we expect that these compounds (if
existing) would have properties similar to MgB$_{2}$. Another
isoelectronic system is lithium borocarbide LiBC\cite{LIBC}, which,
according to our ASA results, is a perfect insulator, so that any
substitution of C by B will lead to metallic behavior. (In general, the
obtained band structure of this system is very similar to that of BN\cite{BN}.)
Experimentally very small conductivity was observed in LiBC\cite{LIBC}.

Addition of one electron to MgB$_{2}$ corresponds to AlB$_{2}$. This
compound also has metallic conductivity but its FS does not have any
cylindrical parts as in MgB$_{2}$. Experimentally no T$_{c}$ has been
observed\cite{ALB2} which makes electron doping an unlikely mechanism to
increase T$_{c}$. Under such circumstances it would be most logical to try
hole doping by replacing Mg by Na or Li. Unfortunately such compounds are
unstable. We are familiar only with one experimental report that NaC$_{2}$%
\cite{NAC2} is a weak SC under pressure and is very unstable. However, it
may be possible to form an alloy (Mg,Na)B$_{2}$ with modest amounts of Na.
It is evident from Fig.~\ref{decomposition} that the hole-type parts of the
FS change dramatically with electron filling.

In summary, MgB$_{2}$ represents the first relatively high T$_{c}$
superconducting compound of boron and has very peculiar bonding
characteristics compared to GIC. We showed that the electronic structure of
MgB$_{2}$ has both similarity with and notable differences to the GICs.
These features provide a basis both for further studies of normal and SC
states of MgB$_{2}$ and for engineering new SC compounds.

We would like to acknowledge useful discussions with S. Bud'ko, P. Canfield,
B. Harmon, K. Ho, I. Mazin and G. Miller. This work was carried out at the
Ames Laboratory, which is operated for the U.S. Department of Energy by Iowa
State University under Contract No. W-7405-82. This work was supported by
the Director for Energy Research, Office of Basic Energy Sciences of the
U.S. Department of Energy. MVS was supported by the Office of Basic Energy
Sciences of the U.S. DOE, Division of Materials Science under contract no.
DE-AC04-94AL85000.

\begin{figure}[tbp]
\caption{Band structures of: (a) top left: primitive (AA stacking) graphite (PG), $%
a=2.456\AA$, $c/a=1.363$; (b) top right: PG boron, $a=3.085\AA$, $c/a=1.142$ (as in MgB$%
_2$); (c) bottom left: LiB$_2$ in MgB$_2$ structure, same $a$ and $c/a$; (d) bottom right: MgB$_2$,
same $a$ and $c/a$. Energy is in eV relative to $E_F$. The order of occupied
bands in the $\protect\Gamma$ point is: $\protect\sigma$ bonding,
$\protect\pi$ bonding, $\protect\sigma$ antibonding (double degenerate).}
\label{manybands}
\end{figure}

\begin{figure}[tbp]
\caption{Pseudocharge density contours obtained in FLMTO. The unit cell is
everywhere that of MgB$_{2}$. Darkness of lines increases with density. (a)
MgB$_{2}$ in (0002) plane passing through B nuclei; (b) MgB$_{2}$ in (1000)
plane passing through Mg nuclei at each corner of the figure. B nuclei
occupy positions (1/3,1/2) and (2/3,1/2) in the plane of the figure. The
integrated charge of the unit cell is 8. (c) (1000) plane, difference in
smoothed density, MgB$_{2}$ minus NaB$_{2}$. The integrated charge of the
unit cell is 1. (d) (1000) plane, difference in smoothed density, MgB$_{2}$
minus PG carbon. The integrated charge of the unit cell is 0. In (c) and
(d), dotted lines show negative values.}
\label{charge}
\end{figure}

\begin{figure}[tbp]
\caption{Rigid band results for MgB$_2$: (a) Total DOS in eV$^{-1}$/cell;
(b) $\protect\sigma_{xx}$ and (c) $\protect\sigma_{zz}$ in
$\rm{Ry}\cdot(a/2\protect\pi)^2$ (for definition of $%
\protect\sigma_{\protect\alpha\protect\beta}$ see text), with contributions
from different parts of the FS. Thick solid lines: total; thin
dashed: 3D electronic part; dotted: 3D hole part; thick dashed: internal
cylinder; thin solid: external cylinder. Energy is in eV relative to $E_F$.}
\label{decomposition}
\end{figure}

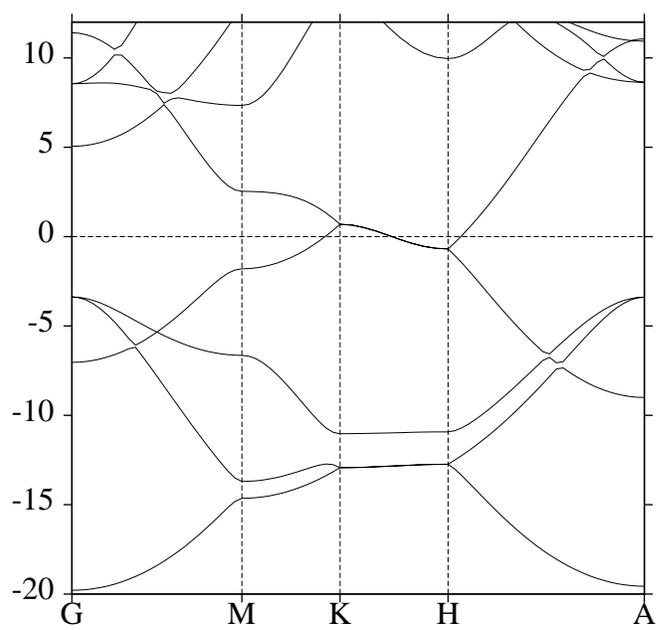
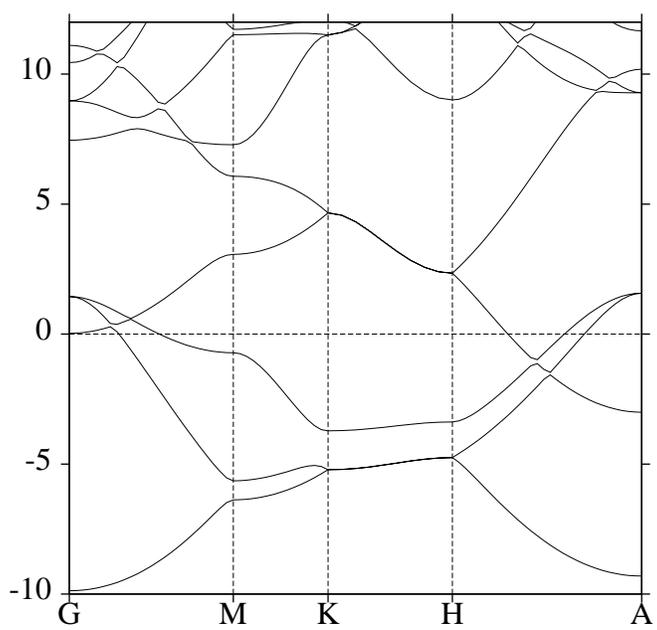
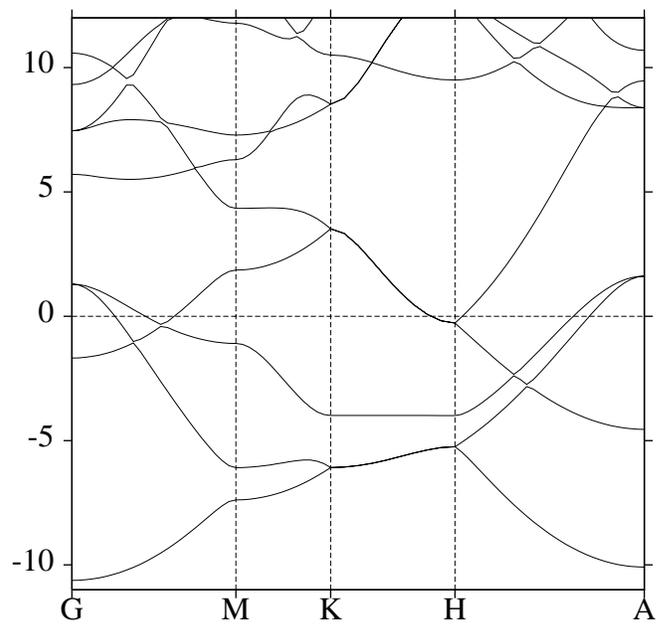
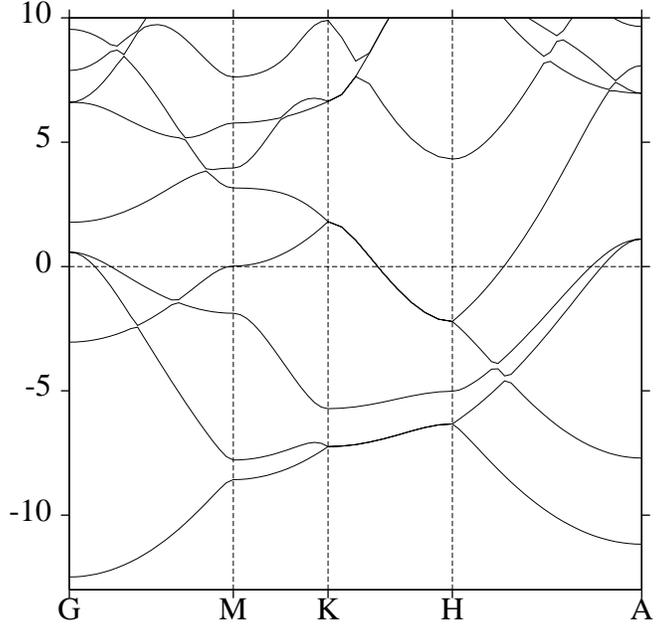
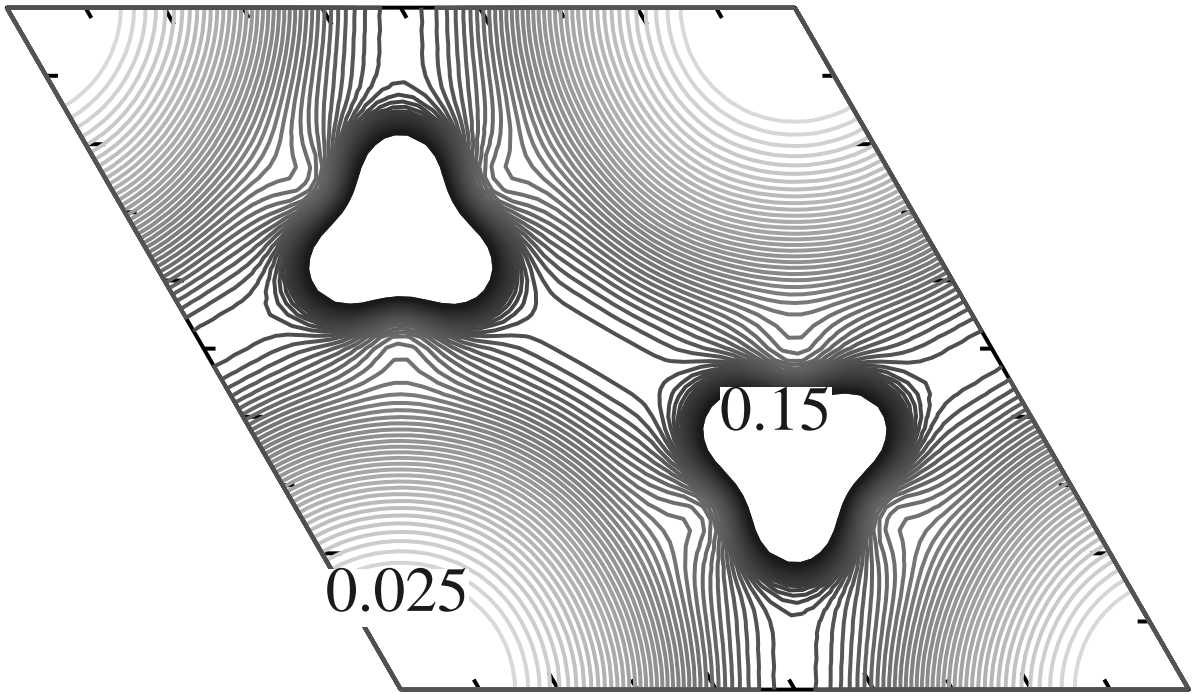
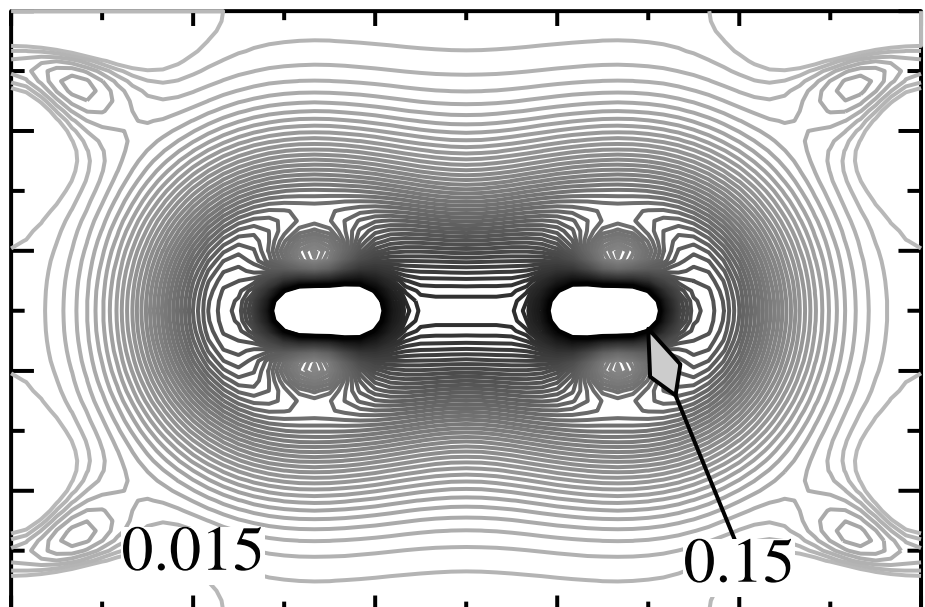
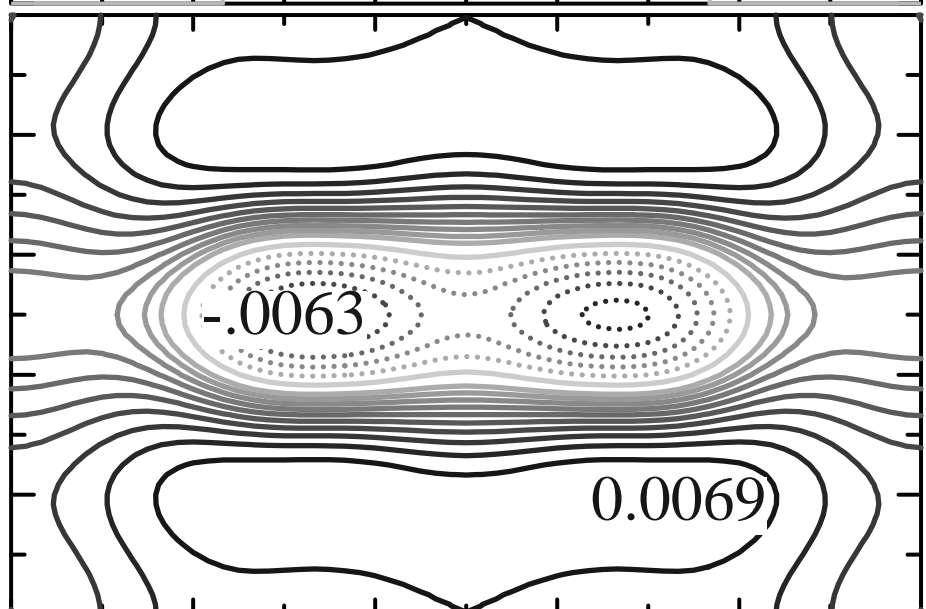
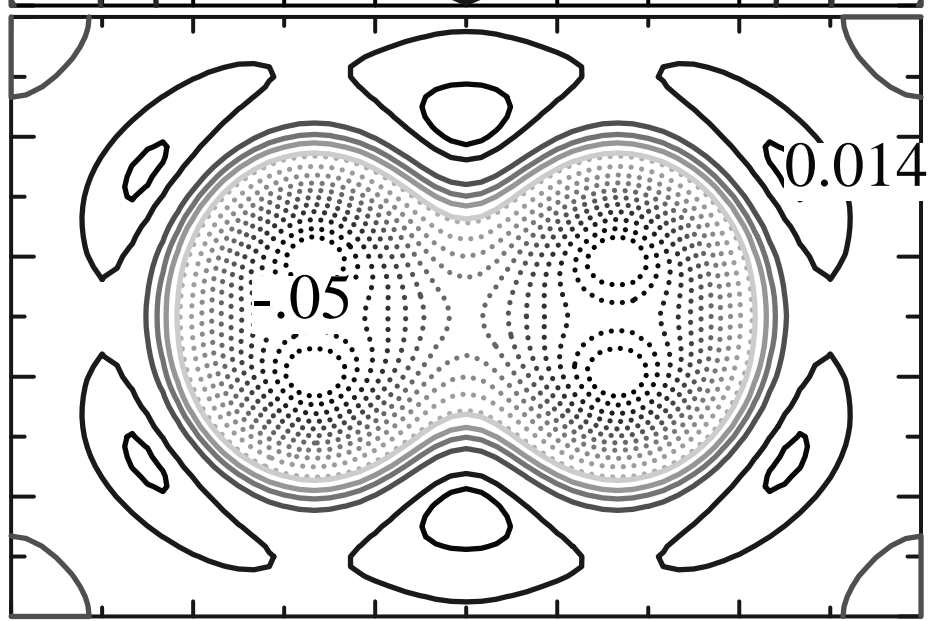
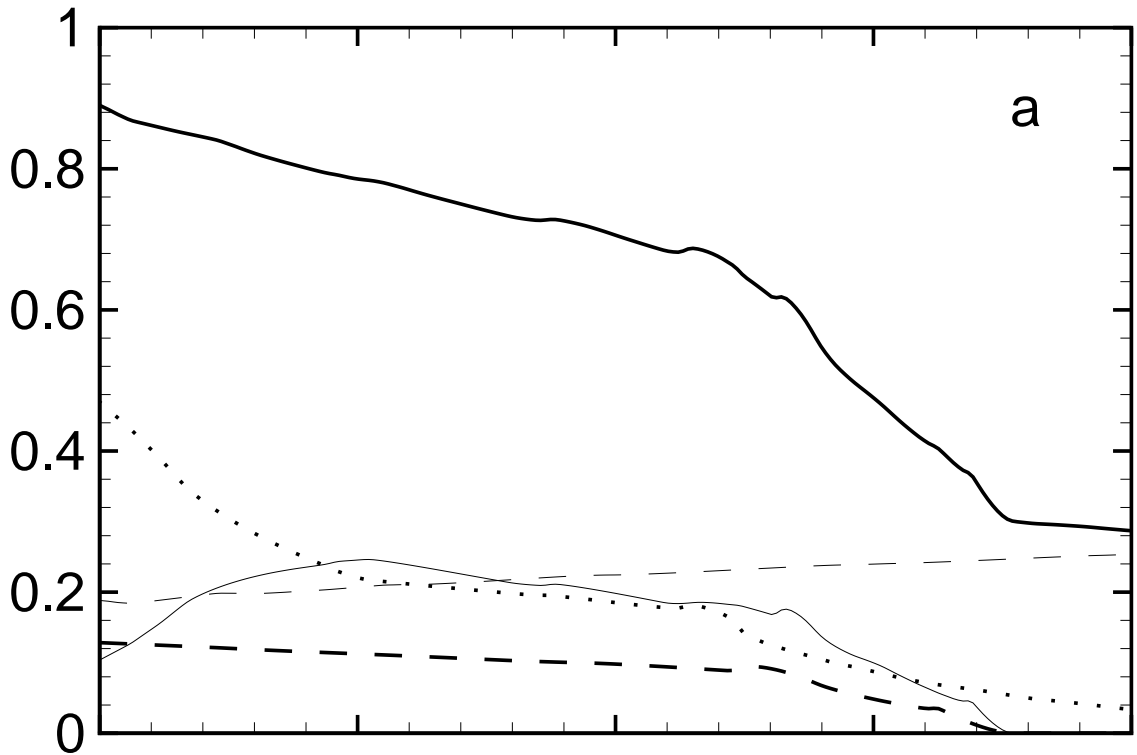
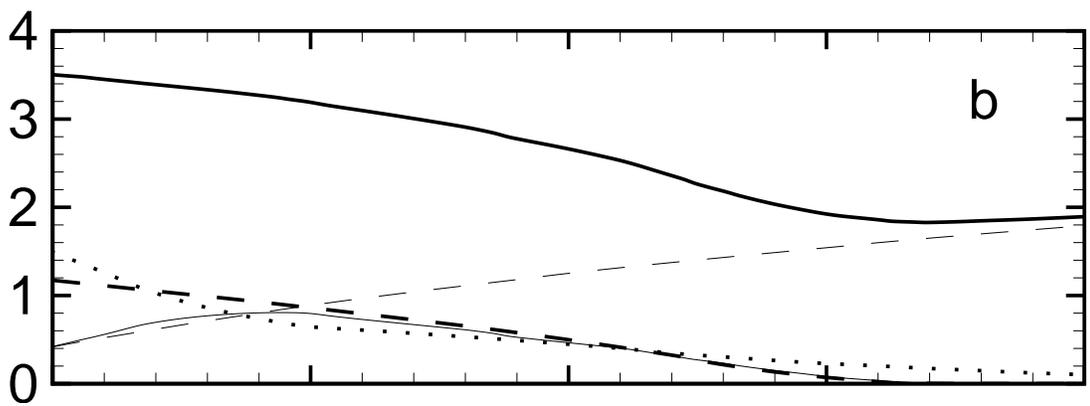
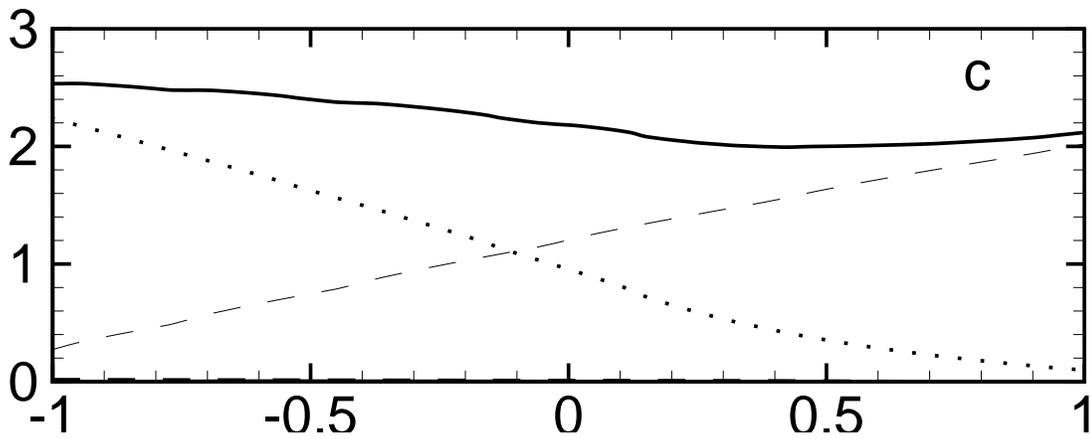
\end{document}